\documentclass[pra,twocolumn,groupedaddress,showpacs,floatfix]{revtex4}

\usepackage{graphics}          
\usepackage{bm}                   
\usepackage{amsmath}         
\usepackage{amsfonts}
\usepackage{amssymb}
\usepackage{latexsym}      
\usepackage{epsfig}   

\begin{document}

\title{Ion-trap quantum computing in the presence of cooling}
\author{Almut Beige}
\affiliation{Max-Planck-Institut f\"ur Quantenoptik, Hans-Kopfermann-Str.~1, 85748 Garching, Germany}
\affiliation{Blackett Laboratory, Imperial College London, Prince Consort Road, London, SW7 2BW, UK\footnote{Present address}} 
\date{\today}

\begin{abstract}
This paper discusses ways to implement two-qubit gate operations for quantum computing with cold trapped ions within one step. The proposed scheme is widely robust against parameter fluctuations and its simplicity might help to increase the number of qubits in present experiments. Basic idea is to use the quantum Zeno effect originating from continuous measurements on a common vibrational mode to realise gate operations with very high fidelities. The gate success rate can, in principle, be arbitrary high but operation times comparable to other schemes can only be obtained by accepting success rates below $80 \%$. 
\end{abstract}
\vspace*{0.2cm}
\noindent

\maketitle

\section{Introduction}

Very recently, considerable progress has been made in the experimental realisation of quantum computing schemes with cold trapped ions. In Innsbruck, the Cirac-Zoller controlled-NOT quantum gate \cite{cz} has been implemented with the help of six concatenated laser pulses individually addressing each of the two ions \cite{nature1}. At the same time, the group in Boulder demonstrated a robust high-fidelity geometric two-qubit phase gate in the laboratory. This was achieved with a sequence of laser pulses and without individual laser addressing of the ions \cite{nature2}. But are these schemes really suitable for quantum computing with many qubits? Finding reliable ways to scale present schemes to many qubits requires simplifications of the experimental setup without decreasing the precision of gate operations. 

\begin{figure}
\begin{minipage}{\columnwidth}
\begin{center}
\resizebox{\columnwidth}{!}{\includegraphics{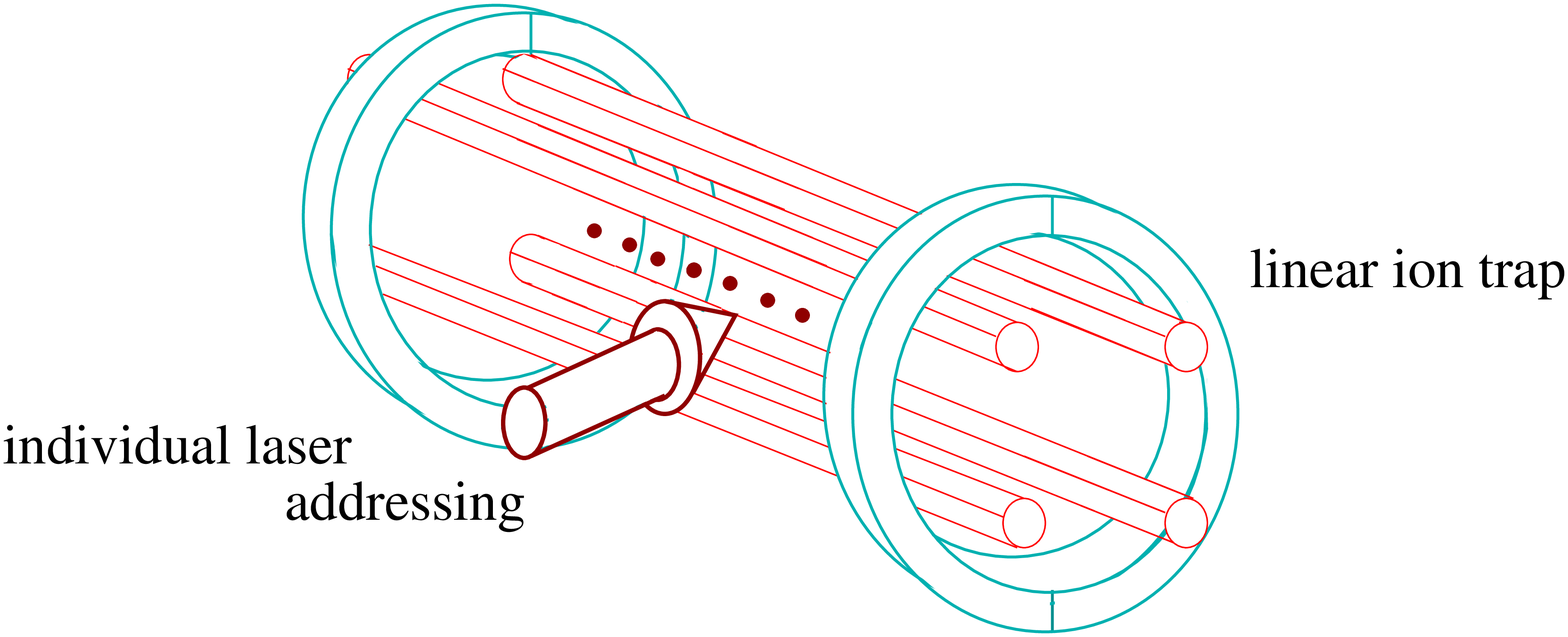}}
\end{center}
\vspace*{-0.7cm}
\caption{Schematic view of the experimental setup. Each qubit is obtained from two different ground states of the same ion. The ions are stored in a linear trap, cooled in the ground state of a common vibrational mode and manipulated with the help of laser fields individually addressing the ions.}\label{trap}
\end{minipage}
\end{figure}

In this paper we present an alternative scheme for the realisation of two-qubit gate operations between cold trapped ions. The ions are stored inside a linear trap (see Figure \ref{trap}) and can be coupled via a common vibrational mode. Each qubit is, as in \cite{simple}, obtained from two different ground states of the same ion and the system remains during the whole computation in the ground state of the common vibrational mode. Quantum gate operations can be implemented within one step by applying different laser fields simultaneously. Depending on the gate operation, this requires individual addressing of the ions by one or two different lasers. Generalising the underlying idea to the many-ion case might enable us in the future to construct efficient schemes for the realisation of many-qubit gates and the generation of highly entangled many-particle states within one step \cite{future}.

A disadvantage of the proposed scheme is that it allows for high gate success rates only in the case of very long gate operation times. If the gate is performed fast, so that its speed becomes comparable to other schemes \cite{nature1}, the gate success rate drops for certain initial states below $80 \%$. However, this disadvantage might be compensated by the fact that the scheme is widely robust against parameter fluctuations, especially against variations of certain Rabi frequencies and detunings. In addition, the lasers for the cooling of the common vibrational mode are in the following applied continuously which should help to decrease decoherence effects due to heating  \cite{eschner,james}. In the setup considered here, there is no need for sympathetic cooling of the ions with the help of a second ion species, as discussed in \cite{mo,morigi}.

The role of cooling in this paper is twofold. On one hand, it decreases the sensitivity of the proposed scheme with respect to heating. On the other hand, the presence of the cooling laser introduces an auxiliary decay channel into the system whose presence restricts the time evolution of the system onto a small subspace such that the implementation of quantum gate operations within one step becomes possible. Photon emissions due to cooling are nevertheless negligible since the system remains continuously in a decoherence-free state \cite{Palma,Zanardi97,Beigenjp}. If heating populates the vibrational mode or gate failure moves the system out of the decoherence-free subspace, then photons are emitted at a high rate \cite{wine79,leib}. This can be detected and the computation can be restarted. For practical purposes, only the fidelity of gate operations must be close to one while the success rate can be significantly smaller. As long as the experiment is repeated if necessary, quantum computers can still be more efficient than conventional computers.

To achieve this, the scheme utilises an environment-induced quantum Zeno effect \cite{misra,behe} originating from the measurements continuously performed by the cooling lasers on the common vibrational mode. This is possible since ground state cooling has the same effect as continuous measurements whether the ions are in the $n=0$ phonon mode or not. The underlying concept of the quantum computing scheme proposed here became known in the past as {\em quantum computing using dissipation} \cite{Letter} and has especially been used to construct simple and precise \cite{Ben} but also very robust \cite{pachos} quantum computing schemes for atom-cavity systems. 

Many schemes for quantum computing with trapped ions have already been proposed. Some of them require cooling of the ions into the ground state of a common vibrational mode \cite{cz}. That this is possible has been demonstrated in several experiments \cite{wine,nature1,nature2,mon}. Other schemes can be implemented with ``hot'' ions \cite{poy,moelm,sarah,jonathan}. For example, the proposal by S{\o}renson and M{\o}lmer \cite{moelm} has been used to entangle up to four ions \cite{sackett} and to observe a violation of Bell's inequality with massive particles \cite{david}. However, most of these schemes are very sensitive to level shifts and require precise control of the ion-phonon interaction. 

The paper is organised as follows. In the next section, we review the basic concept of quantum computing using dissipation. Concrete realisations of two-qubit gate operations with cold trapped ions are discussed in Section III. As examples we consider possible implementations of the CNOT gate, a two-qubit phase gate and the SWAP operation. Section IV analyses the robustness of the proposed quantum computing scheme with respect to fluctuations of detunings and other system parameters. Finally, we conclude our results in Section V.

\section{Quantum computing using dissipation}

An important feature of the proposed quantum computing scheme with cold trapped ions is the auxiliary dissipation channel introduced into the system by continuous laser cooling. Ground state cooling is a process where a detuned laser field transfers the ions into an excited and highly unstable atomic state whenever the phonon mode becomes populated. During spontaneous emission from this level there is a high probability for annihilation of a phonon so decreasing the energy in the vibrational mode. If the system returns into the ground state with $n=0$ phonons, the ions no longer see the cooling laser due to its detuning to the vibrational side band \cite{wine79,eschner,leib} and the time evolution of the system can be described by a Schr\"odinger equation.

Although cooling is applied continuously, spontaneous emission from the ions is negligible.  The reason for this is that the state of the system remains during the whole computation in a decoherence-free subspace (DFS) \cite{Palma,Zanardi97,Beigenjp} and no other states become populated during gate operations. The DFS of a system is a subspace of states whose population does not lead to decoherence. In the presence of cooling, the system of cold trapped ions possesses a non-trivial DFS with respect to spontaneous photon emission. It includes all states with no phonons in the vibrational mode while the ions are in a qubit state or in a highly entangled auxiliary state. 

To calculate the decoherence-free (DF) states of the system and the effective time evolution of the ions during gate operations we use in the following the quantum jump approach \cite{HeWi11}. This method is equivalent to the Monte Carlo wave-function approach \cite{HeWi2} and the quantum trajectory method \cite{HeWi3} and predicts that the initial atomic state $|\psi \rangle$ evolves under the condition of no photon emission as 
\begin{equation} \label{seq} 
|\psi^0(t)\rangle = U_{\rm cond}(t,0) \, |\psi \rangle ~.
\end{equation}
Here $U_{\rm cond}(t,0)$ is the no-photon time evolution operator and corresponds to the non-Hermitian conditional Hamiltonian $H_{\rm cond}$ \cite{HeWi11}. For convenience, $H_{\rm cond}$ has been defined such that 
\begin{equation} \label{P0}
P_0(t,\psi) = \| \, U_{\rm cond} (t,0) \, |\psi \rangle \, \|^2 
\end{equation}
is the probability for no emission in $(0,t)$. 

The basic assumption of the quantum jump approach is that the effect of the environment on the state of a quantum optical system is the same as the effect of rapidly repeated measurements on the free radiation field whether a photon has been emitted or not \cite{schoen}. The non-Hermiticity of the conditional Hamiltonian and the continuous decrease of the amplitude of the state vector $|\psi^0(t) \rangle$ reflect that the observation of no photons reveals information about the system. The longer no photon is emitted the more unlikely it becomes that there is excitation that might cause an emission and the amplitude of a state with a spontaneous decay rate decreases exponentially in time \cite{cook,heger}.

Using the quantum jump approach, a state $|\psi \rangle$ is DF if $P_0(t,\psi) = 1$
for all times $t$ \cite{Beigenjp}. Hence, the DFS of a system is spanned by the eigenvectors of the conditional Hamiltonian with real eigenvalues $\lambda_i$. The eigenvectors $|\lambda_i \rangle$ of $H_{\rm cond}$ are in general non-orthogonal. It is therefore useful to introduce the reciprocal basis vectors $|\lambda^j \rangle$ with $\langle \lambda^j |\lambda_i \rangle = \delta_{ij}$ and to write the conditional Hamiltonian as 
\begin{eqnarray}
H_{\rm cond} &=& \sum_i \lambda_i \, |\lambda_i \rangle \langle \lambda^i|~.
\end{eqnarray}
Suppose that all non-DF states couple strongly to the environment and populating them leads typically to a photon emission within a time $\Delta t$. This time $\Delta t$ is greater than a certain minimal size which can be determined from the quantum jump approach. Provided that the eigenvalues $\lambda_k$ corresponding to 
non-DF states fulfil the condition
 \begin{equation} \label{ass}
{\rm e}^{-{\rm i} \lambda_k \Delta t/\hbar} = 0 ~,
\end{equation}  
the no-photon time evolution operator becomes \cite{remark} 
\begin{equation} \label{ups}
U_{\rm cond}(\Delta t,0) = \sum_{i: |\lambda_i \rangle \in {\rm DFS}} 
{\rm e}^{-{\rm i} \lambda_i \Delta t/\hbar} \, |\lambda_i \rangle \langle \lambda_i|  ~.
\end{equation}
This operator projects every state onto the DFS. The action of the environment over a time $\Delta t$ can therefore be interpreted as a measurement whether the system is DF or not. The probability for no emission in $\Delta t$ equals the probability to be in a DF state. 

The basic idea of {\em quantum computing using dissipation} is to utilise the no-photon time evolution (\ref{ups}) for the implementation of gate operations. As we see below, the continuous measurements caused by the environment on the quantum system lead to a realm of possibilities to induce a DF time evolution between the qubits. The reason for this is that the conditional Hamiltonian $H_{\rm cond}$ can easily be varied by an arbitrary but weak enough interaction without loosing the restriction of the system onto the DFS. As long as the typical time scale of the additional terms in the conditional Hamiltonian is much longer than $\Delta t$ defined by condition (\ref{ass}) \cite{Letter}, the resulting effective time evolution is given by (\ref{ups}) to a very good approximation and can be used to generate a unitary operation between the qubits.

The restriction of the system onto the DFS can intuitively be understood with the help of the quantum Zeno effect \cite{misra,behe}. Within $\Delta t$, a weak interaction can only transfer population proportional to $\Delta t$ out off the DFS. During the next measurement the system is found in a non-DF state with a probability proportional to $\Delta t^2$. Otherwise it is projected back onto the DFS by the time evolution operator (\ref{ups}). The probability to find the system always in a DF state, i.e. $T/\Delta t$ times if $T$ is the gate operation time, goes in the limit of weak interactions, i.e. for $\Delta t/T \to 0$, to one. All transitions out of the DFS are therefore strongly inhibited. 

However, the time evolution inside the DFS is not affected. Using first order perturbation theory with respect to the weak interaction, (\ref{ups}) and the assumption that the system is initially in a DF state, one can show that the conditional time evolution in $\Delta t$ is the same as time evolution obtained from the Hamiltonian
\begin{equation} \label{heff}
H_{\rm eff} = I\!\!P_{\rm DFS} \, H_{\rm cond} \, I\!\!P_{\rm DFS}
\end{equation}
using the same approximations. Here
\begin{equation}
I\!\!P_{\rm DFS} = \sum_{i:|\lambda_i\rangle \in {\rm DFS}} |\lambda_i\rangle \langle \lambda_i|
\end{equation}
is the projector onto the DFS.  The effective Hamiltonian $H_{\rm eff}$ is used in the following to find the appropriate laser configuration for the realisation of certain gate operations.

The reason for the simplicity of the resulting quantum computing schemes is that the DFS of the cold-ion system contains in addition to all ground states, highly entangled states. Through populating these states entanglement between qubits can be created during the effective time evolution even if this would not be possible using the weak interaction alone. In the next section we describe as an example the realisation of universal two-qubits gates within one step by applying different laser fields simultaneously. 

Deviations from the effective time evolution occur if the applied interaction is not weak enough and the system moves in $\Delta t$ slightly out of the DFS. We denote the (non-unitary) correction to the desired effective time evolution in the following by $-A_{\rm corr}(T,0)$ so that
\begin{equation}
U_{\rm cond} (T,0) = U_{\rm eff}(T,0) - A_{\rm corr}(T,0) ~.
\end{equation}
From (\ref{P0}) we then find that the gate success rate $P_0(T,\psi)$ equals
\begin{equation} \label{uhu}
P_0(T,\psi) = 1 - 2 \, {\rm Re} \, \langle \psi | U_{\rm eff}(T,0) A_{\rm corr}(T,0) |\psi \rangle
\end{equation}
while the fidelity of a gate operations under the condition of no photon emission is given by
\begin{equation} \label{F}
F(T,\psi) = {| \langle \psi | U_{\rm eff}(T,0) U_{\rm cond} (T,0) |\psi \rangle |^2 \over P_0(T,\psi)} ~.
\end{equation}
Using Equation (\ref{uhu}) it can be shown that  the fidelity equals one,
\begin{eqnarray}
F(T,\psi) = 1~,
\end{eqnarray}
in first order in the correction $A_{\rm corr}(T,0)$. Dissipation can indeed be used to implement very precise quantum gates with success rates sufficiently close to one. 
 
Let us now estimate the probability of finding the result of a whole computation assuming that each gate can be performed with maximum fidelity but only with a finite success rate $P_0$. The probability of implementing an algorithm of $N$ gates faultlessly is $P_0^N$ and decreases exponentially with $N$. On the other hand, if one always knows whether an algorithm has failed or not, the computation can be repeated until a result is obtained. The probability for not having a result after $M$ runs equals
\begin{equation}
P_{\rm no \, result}= \big(\, 1 - P_0^N \, \big)^M ~.
\end{equation}
For large $N$ this is approximately $\exp (- M P_0^N )$. Many repetitions might be necessary to implement a computation. However, for smaller numbers of $N$ and if $P_0$ is sufficiently close to unity, the failure probability is already nearly negligible for $M \approx N$. For example, if $P_0=95 \%$ and an algorithm with $N=50$ gates is performed, then repeating the computation 50 times yields already a success rate above $98\%$. 

\section{Single laser pulse gates} \label{rough}

\begin{figure}
\begin{minipage}{\columnwidth}
\begin{center}
\resizebox{\columnwidth}{!}{\includegraphics{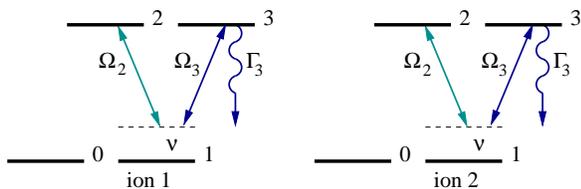}}
\end{center}
\vspace*{-0.4cm}
\caption{Level scheme of the two ions involved in the gate operation. Each qubit is obtained from the ground states $|0\rangle$ and $|1\rangle$ of one ion. In addition a metastable state $|2\rangle$, a rapidly decaying state $|3\rangle$ with decay rate $\Gamma_3$ and two strong laser fields with coupling strength $g_j= {1 \over 2} \, \eta_j \Omega_j$ and detuning $\nu$ are required.} \label{fig1}
\end{minipage}
\end{figure}

In this section we propose concrete realisations of two-qubit quantum gates for cold trapped ions. Each qubit is obtained from two different atomic ground states $|0 \rangle$ and $|1 \rangle$ of the same ion. In addition, a metastable state $|2 \rangle$ and a rapidly decaying level 3 are required, as shown in Figure \ref{fig1}. First, the ions have to be cooled into the ground state of a common vibrational mode.  Two strong laser fields detuned by the frequency $\nu$ of a common vibrational mode should be applied. The laser field coupling to the 1-2 transition  establishes an ``interaction" between the two qubits involved in the gate operation. The laser field driving the 1-3 transition represents the laser cooling setup and can be replaced by any other laser cooling configuration without changing the effective time evolution of the system.

In the following, we denote the spontaneous decay rate of level 3 by $\Gamma_3$ while $b$ and $b^\dagger$ are the annihilation and creation operator of a phonon in the common vibrational mode. The coupling constant of this mode to the atomic 1-$j$ transition equals $g_j \equiv {1 \over 2} \, \eta_j \Omega_j$ where $\Omega_j$ is the Rabi frequency of the applied laser field and $\eta_j$ is the Lamb Dicke parameter depending on the characteristics of the ion trap. Proceeding as in \cite{HeWi11}, one finds that the conditional Hamiltonian within the dipole and the rotating wave approximation and in the interaction picture with respect to the free Hamiltonian equals
\begin{eqnarray} \label{hlaserI}
H_{\rm cond} &=& \sum_{i=1}^2 {\rm i} \hbar \, \big[ \, g_2 \, |1 \rangle_i\langle 2| \, b^\dagger 
+ g_3 \, |1 \rangle_i\langle 3| \, b^\dagger - {\rm h.c.} \, \big] \nonumber \\
&& - \sum_{i=1}^2 {\textstyle{{\rm i} \over 2}} \hbar \Gamma_3 \, |3 \rangle_i\langle 3| ~.
\end{eqnarray}
Here the Lamb-Dicke regime and the condition $\nu \gg \Omega_2$,  $\Omega_3$ has been assumed, as in \cite{cz}. 

The DF states of the two ions are spanned by the eigenstates of the conditional Hamiltonian $H_{\rm cond}$ with no population in the unstable state $|3\rangle$. The DFS of the system therefore contains only superpositions of states with the ions either in the state $|00\rangle$ combined with an arbitrary state of the vibrational mode or both ions in $|01\rangle$, $|10\rangle$, $|11\rangle$ or in the antisymmetric state
\begin{equation} \label{a}
|a\rangle \equiv {\textstyle {1\over \sqrt{2}}} \, \big[ |12 \rangle -|21\rangle \big]
\end{equation}
while the vibrational mode is not populated. The DF states correspond indeed to the eigenvectors of $H_{\rm cond}$ with real eigenvalues. Here these are all zero eigenvalues and the system does not evolve as long as no additional interaction is applied. 

To realise gate operations between the two qubits weak laser fields are required in addition to the strong lasers shown in Figure \ref{fig1}. Let us denote the Rabi frequency of the laser with respect to the $j$-2 transition in ion $i$ by $\Omega_j^{(i)}$ and assume that                                                                          
\begin{equation}
\Omega_j^{(i)} \ll g_2, ~ g_3 ~~ {\rm and} ~~ \Gamma_3~.
\end{equation}
The laser Hamiltonian equals then
\begin{eqnarray} \label{hint}
H_{\rm int} &=& \sum_{i=1,2} \sum_{j=0,1}  {\textstyle {1 \over 2}} \hbar \Omega_j^{(i)} \, |j \rangle_i\langle 2| + {\rm h.c.} 
\end{eqnarray}
If the system is restricted onto the DFS, as predicted in the previous section, then the time evolution of the ions can be predicted with the effective Hamiltonian (\ref{heff}) which yields 
\begin{eqnarray} \label{heff2}
H_{\rm eff} &=& {\textstyle {1 \over 2 \sqrt{2}}} \hbar \, \big[ \, - \Omega_0^{(1)} \, 
|01 \rangle + \Omega_0^{(2)} \, |10 \rangle \nonumber \\
&& + \big( \Omega_1^{(2)} - \Omega_1^{(1)} \big) \, |11 \rangle \, \big] \, \langle a| + {\rm h.c.} 
\end{eqnarray}
The operation time $T$ should be chosen such that all population returns again into the qubit states at the end of each gate operation. 

To analyse the no-photon time evolution of the ions in more detail and to determine the optimal parameter regime for quantum gate implementations the notation
\begin{equation} \label{coeff}
|\psi^0 \rangle \equiv \sum_{n=0}^\infty \sum_{i,j=0}^3 c_{n,ij} \, |n,ij \rangle 
\end{equation}
is introduced. Here $c_{n,ij}$ is the amplitude of the unnormalised state $|\psi^0 \rangle$ with respect to the state with $n$ phonons in the vibrational mode while the ions are in $|ij \rangle$. Under the condition that 
\begin{eqnarray}
\Omega_j^{(i)} \ll g_2, ~ g_3 ~~ {\rm and} ~~ \Gamma_3 ~,
\end{eqnarray}
the coefficients $c_{n,ij}$ of all non-DF states change on a much faster time scale than the coefficients of the DF states. They can therefore be eliminated adiabatically by setting their derivatives equal to zero. Using (\ref{hlaserI}), (\ref{hint}) and the Schr\"odinger equation for the no-photon time evolution we obtain 
\begin{widetext}
\begin{eqnarray} \label{cdgls}
\dot{c}_{0,00} &=& -2 k_1 \, \big[ \, 
\big( \Omega_0^{(1)2}+ \Omega_0^{(2)2} \big) \, c_{0,00} 
+ \Omega_0^{(1)} \Omega_1^{(1)} \, c_{0,10} 
+ \Omega_0^{(2)} \Omega_1^{(2)} \, c_{0,01} \, \big] \nonumber \\
\dot{c}_{0,01} &=& {\textstyle {{\rm i} \over 2 \sqrt{2}}} \, \Omega_0^{(1)} \, c_{0,a}
- k_1 \, \big[ \, 2 \Omega_0^{(2)} \Omega_1^{(2)} \, c_{0,00}
+ \big( 2 \Omega_1^{(2)2}+ \Omega_0^{(1)2} \big) \, c_{0,01} 
+ \Omega_0^{(1)} \Omega_0^{(2)} \, c_{0,10}
+ \Omega_0^{(1)} \big( \Omega_1^{(1)}+ \Omega_1^{(2)} \big) \, c_{0,11} \, \big] 
\nonumber \\ 
\dot{c}_{0,10} &=& -{\textstyle {{\rm i} \over 2 \sqrt{2}}} \, \Omega_0^{(2)} \, c_{0,a}
- k_1 \, \big[ \, 2 \Omega_1^{(1)} \Omega_0^{(1)} \, c_{0,00}
+ \big( 2 \Omega_1^{(1)2}+ \Omega_0^{(2)2} \big) \, c_{0,10} 
+ \Omega_0^{(1)} \Omega_0^{(2)} \, c_{0,01}
+ \Omega_0^{(2)} \big( \Omega_1^{(1)}+ \Omega_1^{(2)} \big) \, c_{0,11} \, \big] 
\nonumber \\
\dot{c}_{0,11} &=& {\textstyle {{\rm i} \over 2 \sqrt{2}}} \, \big[ \, 
\big( \Omega_1^{(1)} - \Omega_1^{(2)} \big) \, c_{0,a}
- k_1 \, \big( \Omega_1^{(1)} + \Omega_1^{(2)} \big) \, 
\big[ \Omega_0^{(1)} \, c_{0,01} + \Omega_0^{(2)} \, c_{0,10} 
+ \big( \Omega_1^{(1)} + \Omega_1^{(2)} \big) \, c_{0,11} \, \big] \nonumber \\
\dot{c}_{0,a} &=& {\textstyle {{\rm i} \over 2 \sqrt{2}}} \, \big[ \,
\Omega_0^{(1)} \, c_{0,01} - \Omega_0^{(2)} \, c_{0,10} 
+ \big( \Omega_1^{(1)} - \Omega_1^{(2)} \big) \, c_{0,11} \, \big]
- k_2 \, \big( \Omega_1^{(1)} - \Omega_1^{(2)} \big)^2 \, c_{0,a} 
\end{eqnarray}
\end{widetext}
with 
\begin{equation} \label{ki}
k_1 \equiv {g_3^2 \over 4 g_2^2 \Gamma_3} ~,~
k_2 \equiv { \Gamma_3 \over 32 g_3^2} + {1 \over 8 \Gamma_3} \, \left[ \, 
{g_2^2 \over g_3^2} + {g_3^2 \over g_2^2} - 1 \, \right] ~.
\end{equation}
As long as the parameters $k_1$ and $k_2$ multiplied with the Rabi frequencies $\Omega_j^{(i)}$ in the effective Hamiltonian (\ref{heff2}) are much smaller than one, the system behaves indeed as predicted by the quantum Zeno effect. 

In the following subsections we discuss in detail three concrete quantum gate implementations, starting with the CNOT gate. This gate, as well as the phase gate, constitutes together with the single-qubit rotation a universal set of quantum gates \cite{michael}. Another gate, which as we will see can easily be performed with a very high fidelity and success rate is the SWAP operation. 

\subsection{CNOT gate}

If ion 1 contains the target qubit and ion 2 provides the control qubit, then the CNOT gate is given by the time evolution operator
\begin{equation}\label{ucnot}
U_{\rm gate} = |00\rangle \langle00| + |01\rangle \langle01| + |10\rangle \langle11| + |11\rangle \langle10| ~.
\end{equation}
The easiest way to realise this gate is when one laser couples with the (real) Rabi frequency $\Omega$ to the 1-2 transition of ion 1 and another one with the same Rabi frequency to the 0-2 transition of ion 2, i.e.
\begin{equation}
\Omega_0^{(2)} = \Omega_1^{(1)} \equiv \Omega ~~ {\rm and}  ~~ \Omega_0^{(1)} =\Omega_1^{(2)}  = 0  ~.
\end{equation}
This parameter choice corresponds to the effective Hamiltonian
\begin{eqnarray} \label{ex1}
H_{\rm eff} &=& {\textstyle {1 \over 2 \sqrt{2}}} \hbar \Omega \, \big[ \, |10 \rangle - |11 \rangle \, \big] \, \langle a|  + {\rm h.c.} 
\end{eqnarray}
If the duration of the laser pulse equals 
\begin{equation}\label{T}
T = 2\pi/\Omega ~, 
\end{equation}
the resulting time evolution is exactly the desired CNOT operation.

From (\ref{cdgls}) and first order perturbation theory we find that the correction of the no-photon time evolution to the effective time evolution of the DF states equals in first order in $k_i \Omega$
\begin{eqnarray} \label{ucond}
A_{\rm corr}(T,0) &=& 2 \Omega^2 k_1 T \, |00 \rangle \langle 00| \nonumber \\
&& + {\textstyle {1 \over 4}} \Omega^2 (7 k_1+k_2)  T \, \big[ |10 \rangle \langle 11| 
+ {\rm h.c.} \big] \nonumber \\
&& + {\textstyle {1 \over 4}} \Omega^2 (5 k_1-k_2) T \, \big[ |10 \rangle \langle 10| + |11 \rangle \langle 11| \big] \nonumber \\
&& - 2\sqrt{2} {\rm i} \, \Omega k_1 \, \big[ |10 \rangle \langle a|  +  |11 \rangle \langle a| +{\rm h.c.} \big] \nonumber \\
&& + \big[ 1- {\textstyle {1 \over 2}} \Omega^2 (k_1+k_2) T \big] \, |a \rangle \langle a| ~.
\end{eqnarray}
This leads to the no-photon probability and gate success rate
\begin{eqnarray} \label{P0gate}
P_0(T,\psi) &=& 1 - 4 \Omega^2 k_1 T \, |c_{0,00}|^2 \nonumber \\
&& - \Omega^2  (5 k_1 - k_2) T \, {\rm Re}  \big[ c_{0,10} c_{0,11}^*\big] \nonumber \\
&& - {\textstyle {1 \over 2}} \Omega^2  (7 k_1 + k_2) T \, \big[ |c_{0,10}|^2 + |c_{0,11}|^2 \big] \nonumber \\
&& - 4 \sqrt{2} \,  \Omega k_1 \, {\rm Im} \big[ c_{0,a} c_{0,10}^* + c_{0,a} c_{0,11}^*\big]  
\end{eqnarray}
which is very close to one as long as $k_1T$ and $k_2T$ are small. Using (\ref{ucond}) one can calculate the gate fidelity under the condition of no photons. As expected, it equals one in first order in $k_i \Omega$.

Let us now determine the parameters that maximise the minimal gate success rate $P_0(T,\psi)$ on average, this means independent from the initial qubit state. From (\ref{P0}) we find that
\begin{equation} 
P_0(T,\psi) \ge \min_i \big\{ \left| \mu_i \right|^2 \big\} 
\end{equation}
if $\mu_i$ are the eigenvalues of the conditional time evolution operator (\ref{ucond}). These are always smaller but close to one. Calculating them we find that up to first order in $k_i$ one has
\begin{equation} 
P_0(T,\psi) \ge 1 - \Omega^2 \max \left\{(k_1+k_2)T, \, 6 k_1 T \right\}  ~.
\end{equation}
Optimising this we find that for a fixed ratio of $\Omega/g_2$ the CNOT gate works most reliable if
\begin{equation} \label{ideal} 
g_{\rm 3 opt} = \sqrt{2} \, g_2 ~~ {\rm and} ~~ \Gamma_{\rm 3 opt} = 2 \sqrt{37} \, g_2 ~.
\end{equation}
For these parameters one has, for all qubit states $|\psi \rangle$,
\begin{equation} \label{bound}
P_{\rm 0 opt}(T,\psi) \ge 1 - 3 \pi \Omega/(\sqrt{37} g_2) ~.
\end{equation}
The success rate of the gate operation assumes its minimum if $| \psi\rangle$ is a superposition of the states $|10\rangle$ and $|11\rangle$. For all other initial qubit states the success rate is even closer or equal to one. 

\begin{figure}
\begin{minipage}{\columnwidth}
\begin{center}
\resizebox{\columnwidth}{!}{\includegraphics{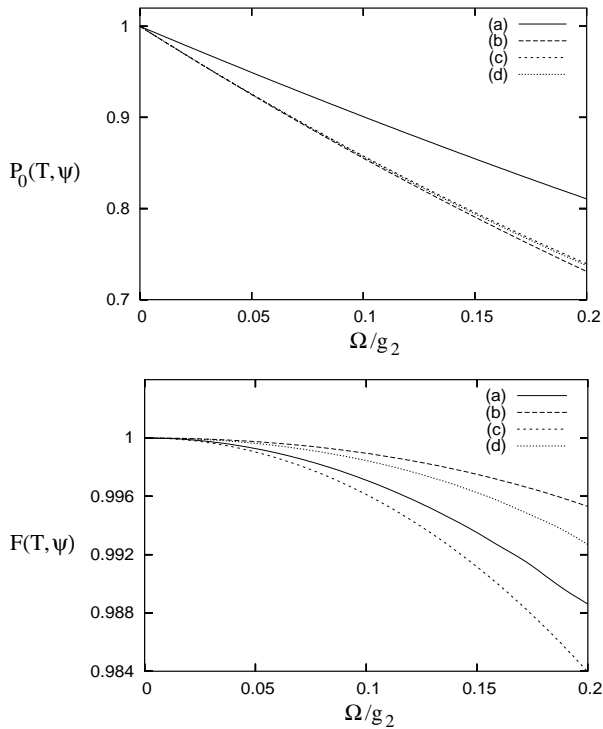}}
\end{center}
\vspace*{-0.5cm}
\caption{Success rate and fidelity under the condition of no photon emission of a single CNOT gate as a function of the Rabi frequency $\Omega$ for $\Gamma_3 = \Gamma_{\rm 3 opt}$, $g_3 = g_{\rm 3 opt}$  and for the initial qubit states $|00 \rangle$ (a), $|10 \rangle$  (b),  $|11 \rangle$ (c) and $[ \, |10\rangle - |11 \rangle \, ]/\sqrt{2}$ (d). The gate success rate and fidelity are always maximal if the ions are initially prepared in $|01 \rangle$.}\label{fig3}
\end{minipage}
\end{figure}

Figure \ref{fig3} results from a numerical solution of the no-photon time evolution given by the sum of the Hamiltonians (\ref{hlaserI}) and (\ref{hint}). The success rate $P_0$ is in good agreement with the theoretical predictions (\ref{bound}). For very small Rabi frequencies, the gate success rate and fidelity is for all initial states close to one. For larger values of $\Omega$, the gate success rate decreases and can for $\Omega = 0.2\, g_2$ be as low as $73 \%$. The gate fidelity is in this case still above $98.4 \%$. The smallest gate success rate is found when the atoms are initially in $|10 \rangle$, $|11 \rangle$ or in a superposition of these two states. Success rates $P_0 > 90 \%$ are achieved as long as $\Omega < 0.07 \,g_2$. The fidelity $F(T,\psi)$ is in this case larger than $99.8 \%$ and the gate duration time $T$ is about $90/g_2$.

\begin{figure}
\begin{minipage}{\columnwidth}
\begin{center}
\resizebox{\columnwidth}{!}{\includegraphics{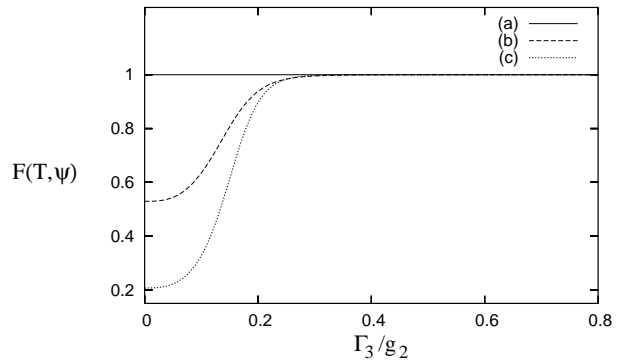}}
\end{center}
\vspace*{-0.5cm}
\caption{Fidelity for a single CNOT gate under the condition of no photon emission as a function of the spontaneous decay rate $\Gamma_3/g_2$ for $g_3=\Gamma_3$, $\Omega=0.01 \, g_2$ and for the initial qubit states $|00 \rangle$ (a), $|10 \rangle$  (b) and $[ \, |10\rangle - |11 \rangle \, ]/\sqrt{2}$ (c). If the ions are initially prepared in $|01 \rangle$, the gate success rate equals one. For $|\psi \rangle = |10 \rangle$ the fidelity is about the same as in graph (c).}\label{fig22}
\end{minipage}
\end{figure}

Finally, we would like to comment on the role of the spontaneous decay rate of level 3 in the scheme. Figure \ref{fig22} shows the fidelity of a single CNOT gate as a function of $\Gamma_3$ and for $g_3=\Gamma_3$. In the chosen parameter regime, the effective damping rate of unwanted population in non-DF states can be shown to equal $g_3^2/\Gamma_3 = \Gamma_3$ to a very good approximation and the effective decay rate of non-DF states increases linearly in $\Gamma_3$. Figure \ref{fig22} confirms that the presence of the auxiliary dissipation channels is crucial for the scheme to work! For small damping rates the mechanism which restricts the time evolution of the system onto the computational subspace fails and the minimum gate fidelity is well below $50 \, \%$. If performed with ions with a larger spontaneous decay rate $\Gamma_3$, the scheme works precisely -- although this parameter should also not be too large since this would introduce an additional fast time scale into the system and the assumptions made to derive the differential equations (\ref{cdgls}) would no longer be valid.

\subsection{Phase gate} \label{phase}

Another two-qubit quantum gate, whose realisation is even simpler than the realisation of the CNOT gate, is the phase gate that changes the sign of the amplitude of the qubit state $|01\rangle$ but leaves all other basis states unchanged. It requires only one weak laser pulse in addition to the laser excitation shown in Figure \ref{fig1} \cite{simple}. This pulse addresses the 0-2 transition in atom 1 and one should choose
\begin{equation}
\Omega_0^{(1)} \equiv \Omega ~~ {\rm and}  ~~ \Omega_0^{(2)} = \Omega_1^{(1)} =\Omega_1^{(2)}  = 0  ~.
\end{equation}
From (\ref{heff2}) we see that this choice of parameters corresponds to the effective Hamiltonian
\begin{eqnarray} \label{hopt}
H_{\rm eff} &=& {\textstyle {1 \over 2 \sqrt{2}}} \, \hbar \Omega \, |01 \rangle \langle a| + {\rm h.c.}
\end{eqnarray}
and the weak laser field should be turned off after a time 
\begin{equation}  \label{reveal0}
T = 2\sqrt{2} \pi/\Omega ~.
\end{equation}
To change the sign of the qubit state $|10 \rangle$ one can proceed in exactly the same way but with  $\Omega_0^{(2)} \equiv \Omega$ and $\Omega_0^{(1)} =  \Omega_1^{(1)} =\Omega_1^{(2)} = 0$.

Using the differential equations (\ref{cdgls}) and first order perturbation theory we find that the correction $A_{\rm corr}(T,0)$ to the effective time evolution equals in first order $k_1\Omega$ 
\begin{eqnarray} \label{ucond2}
A_{\rm corr}(T,0) &=& \sqrt{2} \pi \, k_1 \Omega \, \big[ \, 4 \, |00 \rangle \langle 00| 
- |01 \rangle \langle 01| \, \big] ~.
\end{eqnarray}
This leads to the no-photon probability 
\begin{eqnarray} \label{P0gate2}
P_0(T,\psi) &=& 
1 - 2 \sqrt{2} \pi \, k_1 \Omega \, \big[ \, 4 \, |c_{0,00}|^2 + |c_{0,01}|^2 \, \big] 
\end{eqnarray}
which is very close to unity as long as $k_1\Omega \ll 1$. Calculating the fidelity under the condition of no photon emission (\ref{F}) one finds again that it does not differ from one in first order. 

\begin{figure}
\begin{minipage}{\columnwidth}
\begin{center}
\resizebox{\columnwidth}{!}{\includegraphics{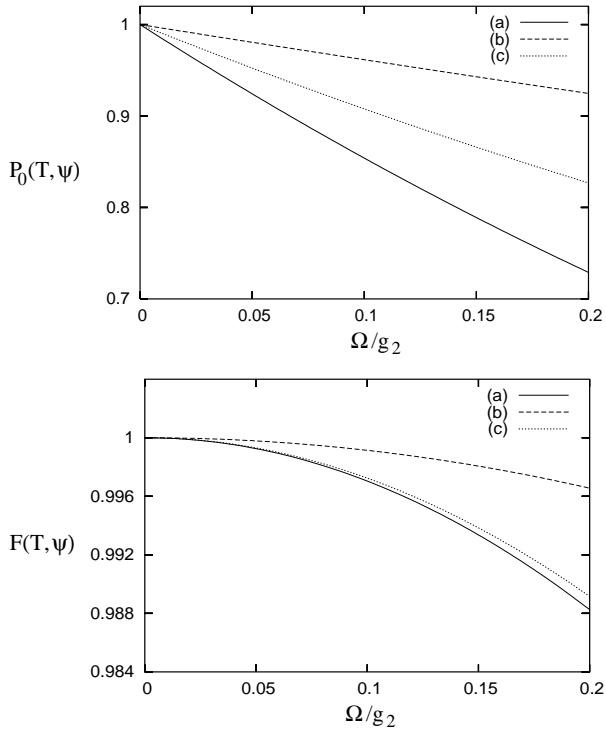}}
\end{center}
\vspace*{-0.5cm}
\caption{Success rate of a single phase gate and fidelity under the condition of no photon emission as a function of the Rabi frequency $\Omega$ for the same parameters $\Gamma_3$ and $g_3$ as in Figure \ref{fig3}.  The initial states of the ions are $|00 \rangle$ (a), $|01 \rangle$  (b) and  $[ \, |00\rangle + |01 \rangle \, ]/\sqrt{2}$ (c).}\label{fig5}
\end{minipage}
\end{figure}

Figure \ref{fig5} results from a numerical solution of the no-photon time evolution of the system for the same parameters as in Figure \ref{fig3}. The success rate and the conditional fidelity of a single phase gate are of similar size as in the previous subsection and the scheme works very well for small Rabi frequencies $\Omega$. As initial states we considered only the the qubit states $|00\rangle$ and $|01\rangle$ since they are the only ones affected by the laser field. If the ions are initially in $|10\rangle$ or $|11\rangle$, the fidelity and success rates of the phase gate are exactly one.

\subsection{SWAP operation}

Another quantum gate that can be implemented relatively easy is the SWAP operation. It transforms the state $|01\rangle$ into $|10\rangle$ and vice versa but leaves $|00\rangle$ and $|11\rangle$ unchanged. The realisation of this gate does not require individual laser addressing of the ions and one should choose
\begin{equation}
\Omega_0^{(1)} = \Omega_0^{(2)} \equiv \Omega ~~ {\rm and}  ~~ \Omega_1^{(1)} =\Omega_1^{(2)}  = 0 ~.
\end{equation}
For these Rabi frequencies the effective Hamiltonian (\ref{heff2}) becomes
\begin{eqnarray} \label{hopt2}
H_{\rm eff} &=& {\textstyle {1 \over 2 \sqrt{2}}} \, \hbar \Omega \, \big[ \, - |01 \rangle + |10 \rangle \, \big] 
\, \langle a| + {\rm h.c.}
\end{eqnarray}
and the duration of the weak laser pulse should equal 
\begin{equation}
T = 2 \pi/\Omega ~.
\end{equation}
The SWAP gate can be very useful. It exchanges the states of two qubits without that the corresponding ions have to swap their places physically. 

Proceeding as in the previous subsection, the correction to the desired effective time evolution reads
\begin{eqnarray} \label{ucorr}
A_{\rm corr}(T,0) &=& 2 \pi \, k_1 \Omega \, \big[ \, 4 \, |00 \rangle \langle 00| 
+ |01 \rangle \langle 01| + |10 \rangle \langle 10| \nonumber \\
&& + |01 \rangle \langle 10| + |10 \rangle \langle 01| \, \big] 
\otimes |0\rangle_{\rm vib} \langle 0| 
\end{eqnarray}
and the gate success rate becomes
\begin{eqnarray} \label{P0gate3}
P_0(T,\psi) &=&  1 - 4 \pi \, k_1 \Omega \, \big[ \, 4 \, |c_{0,00}|^2 + |c_{0,01}|^2 + |c_{0,10}|^2 \nonumber \\
&& + 2 \, {\rm Re} \, \big(c_{0,01} c_{0,10}^*\big) \, \big] ~.
\end{eqnarray}
Again, the fidelity under the condition of no photon emission can be shown to equal one in first order $k_1\Omega$. A comparison with (\ref{P0gate}) shows that the success rate of each SWAP gate is of similar size as for the phase gate discussed in the previous subsection. 

\subsection{Dissipation-assisted quantum gates}

A closer look at the differential equations (\ref{cdgls}) reveals that for fixed gate operation times relatively small corrections to the effective time evolution (\ref{heff2}) are obtained if 
\begin{equation} \label{huhu}
\Omega_1^{(1)}-\Omega_1^{(2)}=0 ~~ {\rm and} ~~ \Gamma_3 \gg g_3~. 
\end{equation}
Then all terms proportional to $k_2$ vanish in (\ref{cdgls}) and the terms proportional to $k_1$ become relatively small. The corresponding gates are relatively easy to realise because they require only one weak laser field applied to the 0-2 transition in each atom. The effective Hamiltonian is in this case given by
\begin{eqnarray} \label{hopt3}
H_{\rm eff} &=& {\textstyle {1 \over 2 \sqrt{2}}} \hbar \, \big[ \, - \Omega_0^{(1)} \, 
|01 \rangle + \Omega_0^{(2)} \, |10 \rangle \, \big] \, \langle a| + {\rm h.c.} \nonumber \\ &&
\end{eqnarray}
and can be realised with $\Omega_1^{(1)}=\Omega_1^{(2)}=0$. The above described phase gate and SWAP operation are examples for quantum gates that can be implemented in this way and they are the only gates for whom the presence of the auxiliary dissipation channel is not crucial for the scheme to work.  

\begin{figure}
\begin{minipage}{\columnwidth}
\begin{center}
\resizebox{\columnwidth}{!}{\includegraphics{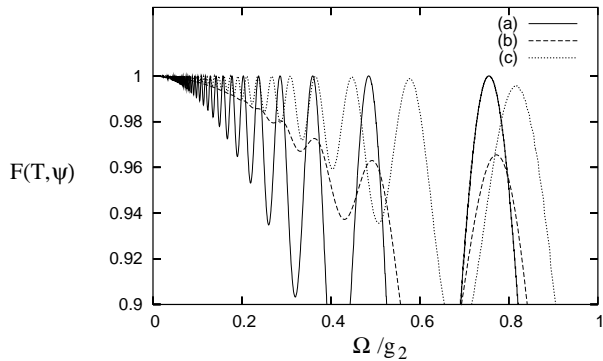}}
\end{center}
\vspace*{-0.5cm}
\caption{Fidelity for a single phase gate under the condition of no photon emission as a function of the Rabi frequency $\Omega$ if the ions are initially prepared in the state $|00 \rangle$ and for $g_3=\Gamma_3 =0$ (a), for $g_3=g_2$ and $\Gamma_3 =10 \, g_2$ (b) and for $g_3=g_2$ and $\Gamma_3 =0$ (c). The curves give a lower bound for the fidelity of the prepared state independent from the initial state.}\label{fig6}
\end{minipage}
\end{figure}

Remarkable about the effective time evolution (\ref{hopt3}) is that cooling plays indeed a different role for its realisation than in the case of the CNOT gate which only works because of dissipation. Suppose the quantum Zeno effect provides the basic mechanisms that restricts the system onto the DFS. Then one would expect that the quantum gate works the better if one increases the value of $g_3^2/\Gamma_3$ since this number relates directly to the spontaneous decay rate of non-DF states. Instead it can be shown that the phase gate and the SWAP operation  work the better the smaller $g_3^2/\Gamma_3$ since this implies smaller values for $k_1$. Both gates can even be implemented in the absence of cooling for $g_3=\Gamma_3=0$ \cite{simple}.

The mechanism which yields the effective Hamiltonian (\ref{hopt3}) in this subsection is adiabaticity resulting from the presence of two different times scales in the system. One is given by the weak Rabi frequencies $\Omega_0^{(i)}$ and the other one by the coupling constant $g_2$. To guarantee a success rate close to one, quantum gates based on adiabatic passages are as slow as gates based on a quantum Zeno effect. Advantages arise when the system is operated outside the adiabatic regime with gate operation times of about the same size as $1/g_2$. In this regime, cooling has the same effect as error detection measurements. Under the condition of no photon emission, the system behaves as predicted by adiabaticity. The presence of the auxiliary dissipation channel exponentially damps away the population that accumulates due to non-adiabaticity in unwanted states. The resulting process can be called a {\em dissipation-assisted adiabatic passage} \cite{simple}. The price one pays for the speed up of gate operations is a decrease of the gate success rate $P_0$. 

Figure \ref{fig6} shows the fidelity of a single phase gate under the condition of no photon emission for a wider range of Rabi frequencies $\Omega$ than assumed in Figures \ref{fig3} and \ref{fig5}. The fidelity of the finally obtained state is for $g_3=\Gamma_3=0$ very close to one as long as $\Omega$ is much smaller than $g_2$  (see Figure \ref{fig6}(a)). Note that the same fidelity close to one can be achieved for higher Rabi frequencies, and therefore also for shorter gate operation times, in the presence of dissipation (see Figure \ref{fig6}(b)). Choosing $g_3=g_2$ and $\Gamma_3=0$ improves the precision of the performed gate operation even further, corresponds to the gate success rate $P_0 \equiv 1$ but does not provide cooling of the ions during gate operations (see Figure \ref{fig6}(c)).

\section{Robustness against parameter fluctuations}

In this Section we discuss in detail the robustness of the proposed quantum computing scheme with respect to parameter fluctuations and analyse the effect of variations of Rabi frequencies and the presence of detunings. The independence of the scheme from the precise value of most experimental parameters is a big advantage of the proposed quantum computing scheme and applies to quantum computing schemes using dissipation in general. 

\subsection{Effect of detunings}

In the following we denote the detuning of the laser field that establishes the coupling of the ions to the common vibrational mode by $\Delta_2$. The reason for the independence of the no-photon time evolution of the system from this parameter arises from the different role of the ion-phonon interaction in the present scheme compared to other quantum computing schemes with cold trapped ions. Here the interaction does {\em not} cause a transfer of excitation into a common vibrational mode. Instead, its presence introduces a fast time scale into the system which prevents transitions into unwanted states. During gate operations, the phonon mode becomes only virtually populated.

\begin{figure}
\begin{minipage}{\columnwidth}
\begin{center}
\resizebox{\columnwidth}{!}{\includegraphics{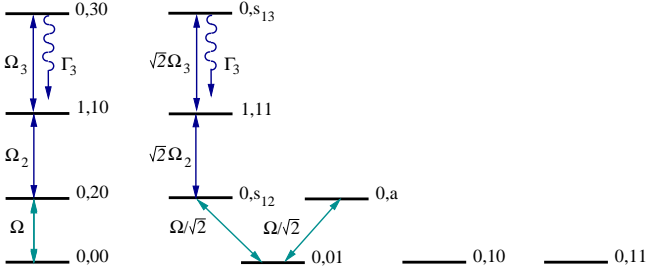}}
\end{center}
\vspace*{-0.2cm}
\caption{Level scheme showing {\em all} transitions and level involved in the realisation of the two-qubit phase gate.}\label{level}
\end{minipage}
\end{figure}

To visualise the role of the vibrational mode in the scheme it is useful to have a look at all levels and transitions involved in the time evolution of the system. As an example we survey in the following the phase gate which has been introduced in Section \ref{phase}. Its level scheme is shown in Figure \ref{level} and is much simpler than  the level configurations for the CNOT gate and the SWAP operation. Still, the independence of detunings described here applies to single laser pulse gates with cold trapped ions in general and can be shown for all other gates in the same way.

If initially only qubit states are populated, then the time evolution of the system during a phase gate implementation involves only the eleven levels shown in Figure \ref{level}. Let us define the interaction-free  Hamiltonian $H_0$ as 
\begin{eqnarray} \label{hl}
H_0 &=& H_{\rm ions} + H_{\rm phonons} - \hbar \Delta_2  \big[ \, |1,10 \rangle \langle 1,10|  \nonumber \\
&& +  |1,11 \rangle \langle 1,11| + |0,30 \rangle \langle 0,30| +  |0,s_{13} \rangle \langle 0,s_{13}| \, \big]  ~, \nonumber \\ &&
\end{eqnarray}
where $\Delta_2$ is the detuning of the strong laser field driving the 1-2 transition of each ion and
\begin{equation} \label{s}
|s_{1j} \rangle \equiv {\textstyle {1\over \sqrt{2}}} \, \big[ |1j \rangle + |j1\rangle \big] ~.
\end{equation}
In the interaction picture with respect to this Hamiltonian, the Hamiltonian describing the no-photon time evolution of the system becomes
\begin{eqnarray} \label{hl222}
H_{\rm cond} \nonumber 
&=& {\rm i} \hbar g_2 \, \big[ \, |0,20 \rangle \langle 1,10| + \sqrt{2} \, |0,s_{12} \rangle \langle 1,11| \, \big] \nonumber \\
&& + {\rm i} \hbar g_3 \, \big[ \, |1,10 \rangle \langle 0,30| + \sqrt{2} \, |1,11 \rangle \langle 0,s_{13}| \, \big] \nonumber \\
&& + {\textstyle {1 \over 2}} \hbar \Omega \, \big[ \,  |0,00 \rangle \langle 0,20| 
+ {\textstyle {1 \over \sqrt{2}}} \,  |0,01 \rangle \langle 0,s_{12}| \nonumber \\
&& + {\textstyle {1 \over \sqrt{2}}} \,  |0,01 \rangle \langle 0,a| \, \big]  + {\rm h.c.} + \hbar \Delta_2 \, \big[  \, |1,10 \rangle \langle 1,10| \nonumber \\
&& + |1,11 \rangle \langle 1,11| \, \big] - \hbar \, \big( {\textstyle{{\rm i} \over 2}} \Gamma_3 - \Delta_2 \big) \, 
\big[ \, |0,30 \rangle \langle 0,30| \nonumber \\
&& + |0,s_{13} \rangle \langle 0,s_{13}| \, \big] ~. 
\end{eqnarray}
One can easily check that the effective Hamiltonian (\ref{heff}) of the system does not depend on the detuning $\Delta_2$. In a first approximation, the time evolution of the system does not change, even if the detuning $\Delta_2$ is about the same size as $g_2$.

\begin{figure}
\begin{minipage}{\columnwidth}
\begin{center}
\resizebox{\columnwidth}{!}{\includegraphics{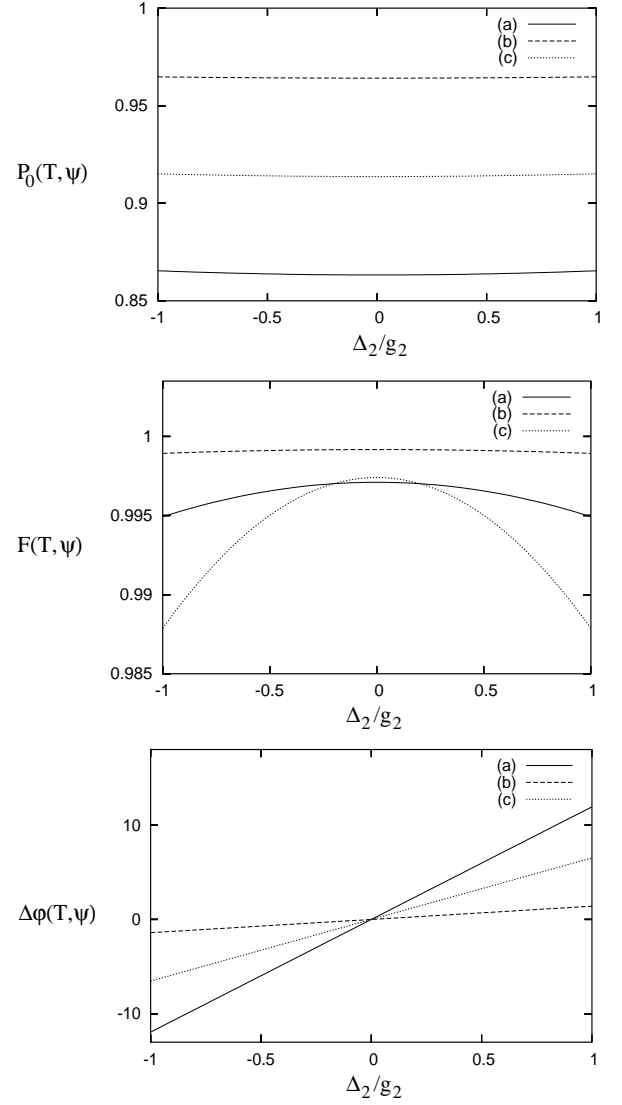}}
\end{center}
\vspace*{-0.5cm}
\caption{Success rate of a single phase gate and fidelity under the condition of no photon emission as a function of the detuning $\Delta_2$ of the ion-coupling laser for the same parameters and initial states as in Figure \ref{fig5}. The last graph gives the size of the phase error $\Delta \varphi$ of the amplitude of the final state in degrees.}\label{delta}
\end{minipage}
\end{figure}

This is in good agreement with the numerical results shown in Figure \ref{delta} which has been obtained by solving the no-photon time evolution of the system with the Hamiltonian (\ref{hl}). As expected, the presence of the detuning $\Delta_2$ has only a small effect on the success rate of the scheme. The fidelity of the gate operation under the condition of no photon emission decreases less than one percent even if $\Delta_2$ becomes as big as the ion-phonon coupling constant $g_2$. However, the amplitude of the finally prepared state collects a phase factor in the presence of detuning. Depending on the initial qubit state, this phase can differ up to $12^\circ$ from the desired phase (see Figure \ref{delta}). 

Analogously, it can be shown that detunings in the cooling laser setup affect the effective time evolution of the system only slightly. Detunings of the weak laser field with Rabi frequency $\Omega$ have to be taken into account because they lead to an additional term in the effective Hamiltonian (\ref{heff}). 

\subsection{Fluctuations of Rabi frequencies} \label{fun}

Let us now assume again zero detunings and discuss the effect of fluctuations of Rabi frequencies on the desired time evolution of the system. The analysis in Section \ref{rough} showed already that the performance of the proposed quantum gate operations does not depend on the size of the ion-phonon coupling constants $g_2$ and $g_3$ and the decay rate $\Gamma_3$ as long as these rates are big enough. In the formalism this is reflected by the fact that the effective Hamiltonians (\ref{ex1}), (\ref{hopt}) and (\ref{hopt2}) depend only on $\Omega$. In this subsection we show that the scheme is also widely robust against fluctuations of the weak Rabi frequency $\Omega$. 

The reason for this robustness is that the effective Hamiltonian can be written as $H_{\rm eff} = \Omega M$ where $M$ is a time independent operator. This is the case for all three quantum gates discussed in the previous Section. Let us assume that the Rabi frequency $\Omega = \Omega(t)$. The effective time evolution operator of the system is then given by
\begin{eqnarray}
U_{\rm eff}(T,0) &=& \exp \left[ \, -{{\rm i} \over \hbar} \int_0^T {\rm d}t \, \Omega(t) M \, \right]
\end{eqnarray}
and depends only on the integral $\int_0^T {\rm d} t \, \Omega(t)$. High fidelities require therefore only precise control of this time integral. With respect to fluctuations of system parameters, quantum computing schemes using dissipation share some features with schemes that employ holonomies or geometrical time evolutions \cite{pachoszanardi} or similar dynamical processes to implement quantum gate operations.

Finally, we would like to comment on the effect of shaped laser pulses on the quantum gate performance. As an example we consider again the phase gate and assume 
\begin{equation} \label{xxx}
\Omega(t) = 2\Omega_0 \, \sin^2 \Big[\, {\textstyle {{1 \over 2\sqrt{2}}}} \Omega_0 t\, \Big] 
\end{equation}
while the gate operation time should be chosen as
\begin{equation} \label{reveal}
T= 2\sqrt{2} \pi/\Omega_0 ~.
\end{equation}
The integral $\int_0^T {\rm d}t \, \Omega(t)$ is then the same as in Subsection \ref{phase} and the Rabi frequency $\Omega$ equals zero at the end of the gate operation. Motivation for assuming a time dependence as in (\ref{xxx}) is to minimise the corrections to the desired time evolution and to increase the fidelity of the finally prepared state without increasing the gate operation time. 

\begin{figure}
\begin{minipage}{\columnwidth}
\begin{center}
\resizebox{\columnwidth}{!}{\includegraphics{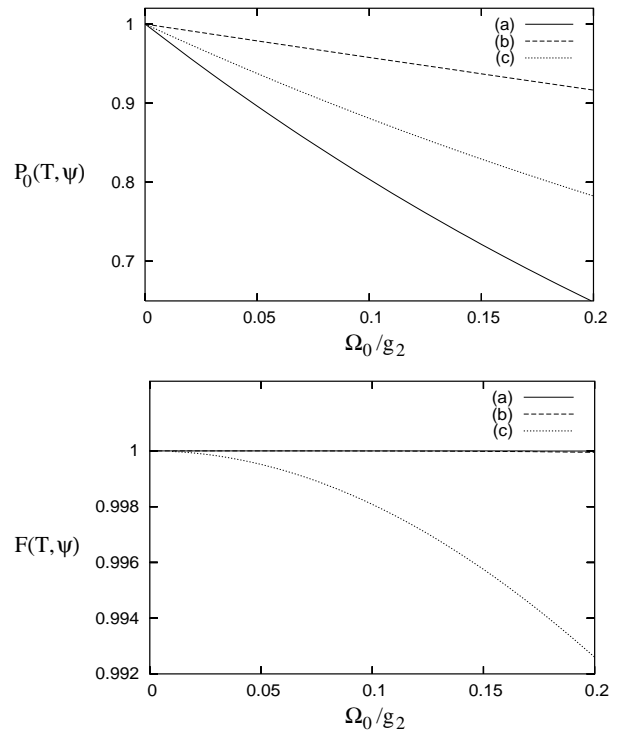}}
\end{center}
\vspace*{-0.5cm}
\caption{Success rate of a single phase gate and fidelity under the condition of no photon emission as a function of the Rabi frequency $\Omega_0$ (defined in (\ref{xxx})) for the same parameters and initial states as in Figure \ref{fig5}.}\label{final}
\end{minipage}
\end{figure}

Figure \ref{final} results from a numerical simulation of the no-photon time evolution of the system and shows that $F(T,\psi)$ can be improved even further. The gate fidelity is now for many initial states closer to one and for all initial states above $99.2 \%$, even if $\Omega_0=0.2 \, g_2$. A comparison of Figure \ref{final} with Figure \ref{fig5} reveals that the price one has to pay for this improvement of the precision of the scheme is a reduction of the gate success rate $P_0(T,\psi)$ for certain initial qubit states by a few percent. (Note that the gate operation times $T$ are in both Figures the same when $\Omega$ and $\Omega_0$ are the same.) Similar results can be derived for other single-laser pulse quantum gates with cold trapped ions.

\section{Conclusions}

We discussed the possibility to implement precise quantum gates between cold trapped ions in the presence of cooling of a common vibrational mode within one step. That the proposed quantum computing scheme is feasible with present technology has recently been demonstrated in an experiment in Innsbruck \cite{nature1}. The setup for the realisation of the Cirac-Zoller CNOT gate with Calcium ions reported in \cite{nature1} is very similar to the setup needed for the implementation of the two-qubit phase gate and the SWAP operation presented in this paper. Both quantum computing schemes require cooling of the ions to the vibrational ground state and a strong laser field to establish a coupling between the ions. In addition only one laser pulse is needed. To implement the phase gate, this laser has to excite one of the ions resonantly. 

Because of its simplicity and robustness, the proposed scheme might help to increase the number of qubits in present quantum computing experiments. One of its advantages is that it offers a great variety of ways to implement gate operations, which can all be performed within one step. Laser cooling is not only possible between computations but should be applied continuously. An additional ion species for sympathetic cooling, as proposed in \cite{morigi}, is not required. Although the system remains continuously in the ground state of a common vibrational mode, the presence of the cooling laser setup decreases the probability for heating even if no photons are actually scattered from the ions \cite{eschner}.

Another major benefit of the proposed quantum computing scheme is its independence from fluctuations of most experimental parameters. To guarantee a high precision of gate operations only one of the applied laser fields has to be controlled relatively well. Its Rabi frequency integrated over the gate operation time has to equal a certain value.  All other Rabi frequencies, $\Omega_2$ and $\Omega_3$, do not enter the effective time evolution of the system and the scheme does not depend on their concrete value as long as they are big enough. The proposed setup can also tolerate a detuning of the laser field which establishes the coupling of the ions to the vibrational mode, even if this detuning is of the same size as the ion-phonon coupling constant $g_2$.

Error detection is already included in the scheme \cite{simple}. If a gate operation fails or heating leads to the population of higher phonon modes, then photons are emitted at a high rate. This can easily be detected and the whole computation can be restarted if necessary. In this way, the system is protected against certain kinds of errors. As long as the gate success rate is sufficiently close to one, quantum computing can still be more efficient than performing the same computation on a conventional computer.

A disadvantage of the presented quantum computing scheme is that its gate operation times are longer than in many other setups. For example, the implementation of a single two-qubit phase gate requires the time $T=2 \sqrt{2} \pi/\Omega$ where $\Omega$ can be as large as $0.1$ times the ion-phonon coupling constant $g_2$. Under the condition of no photon emission, this guarantees a fidelity above $99.6 \, \%$ and a gate success rate above $85 \, \%$, independent of the initial state of the ions (see Figure \ref{fig5}). Higher fidelities and success rates are obtained for smaller Rabi frequencies and longer gate operation times. \\
  
{\em Acknowledgement.} A.B. would like to thank Ferdinand Schmidt-Kaler, Wolfgang Lange and David J.
Wineland for helpful and interesting comments. This work was supported by the Royal Society in form of a University Research Fellowship and by the EPSRC and the European Union in part.

\narrowtext


\begin{references}
\bibitem{remark}
The eigenvectors $|\lambda_i \rangle$ of $H_{\rm cond}$ which correspond to DF states are orthogonal to each other and one has $|\lambda^i \rangle = |\lambda_i \rangle$.
\end{references}
\end{document}